\documentclass[%
 reprint,
showpacs,
 amsmath,amssymb,
 aps,
prb,
]{revtex4-1}

\usepackage{graphicx}
\usepackage{dcolumn}
\usepackage{bm}
\usepackage{multirow}
\usepackage{tikz}

\begin{document}

\preprint{APS/123-QED}

\title{The importance of anisotropic Coulomb interactions and exchange to the band gap and antiferromagnetism of $\beta$-MnO$_2$ from DFT+U}

\author{D.A. Tompsett}
 \email{dt331@bath.ac.uk}
\author{D.S. Middlemiss}%
\author{M.S. Islam}%
\affiliation{Department of Chemistry, University of Bath, Bath BA2 7AY, UK.
}%
\affiliation{Department of Chemistry, University of Cambridge, Cambridge CB2 1EW, UK.}

\date{\today}

\begin{abstract}
First principles density functional theory (DFT) is used to investigate the electronic structure of $\beta$-MnO$_2$. From collinear spin polarized calculations we find that DFT+U$_{\mathrm{Eff}}$ predicts a gapless ferromagnet in contrast with experiment which indicates an insulating antiferromagnet. The inclusion of anisotropic Coulomb and exchange interactions in the DFT+U approach, defining $U$ and $J$ explicitly, corrects these errors and leads to an antiferromagnetic ground state with a fundamental gap of 0.8 eV consistent with low temperature experiments. To our knowledge, this work on $\beta$-MnO$_2$ represents the first demonstration of a case in which the application of fully anisotropic interactions in DFT+U determines the magnetic order and consequent band gap, while the more commonly used effective U approach fails. Such effects are argued to be of importance in many insulating materials. The mechanism leading to an increase in band gap due to anisotropic interactions is highlighted by analytical calculation of DFT+U \textit{d}-orbital eigenvalues obtained within a Kanamori-type model. Magnetic coupling constants obtained by the fitting of a Heisenberg spin Hamiltonian to the energies of a range of magnetic states assist in rationalizing the finding that anisotropic interactions enhance the stability of the experimentally observed helical antiferromagnetic order. The plane wave PAW method yields poorer results for the exchange couplings than full-potential LAPW calculations. Finally, we compare the DFT+U results with exchange couplings obtained from hybrid functionals. It is argued that anisotropic interactions should be included in DFT+U if the results are to be properly compared with those from hybrid functionals.
\end{abstract}

\maketitle


\section{\label{sec:introduction}Introduction}
Density functional theory has been used with success in understanding the properties of many materials. However, in the class of transition metal oxides the performance of traditional DFT methods, typically based upon local spin density or generalized gradient (GGA) approximations, is often found to be inadequate. For instance, these methods predict metallic states for many insulating transition-metal oxides\cite{Anisimov1, Franchini1}. This is understood to be due to the presence of strong on-site Hubbard U interactions resulting from the electronic correlations in open $d$ and $f$ shells. The DFT+U approach has been introduced\cite{Anisimov1} specifically to overcome such deficiencies. This approach incorporates an enhanced treatment of Coulomb and exchange interactions in self-consistent DFT, improving the description of such materials in many cases. The present study of $\beta$-MnO$_2$ makes the further point that an accurate treatment of fully anisotropic Coulomb and exchange interactions may be of vital importance in application of the DFT+U method, proceeding beyond the conceptually simpler `effective U' approach that is often applied.  

\begin{figure}
\includegraphics[scale=0.13,angle=0]{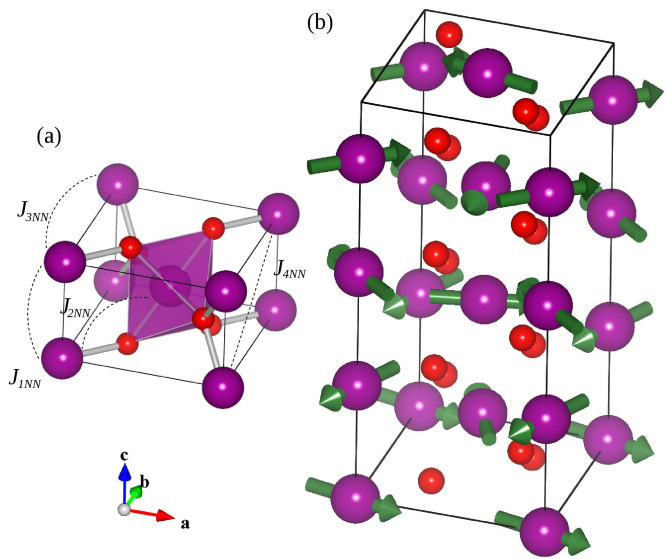}
  \caption{(Color online) (a) The structural unit cell of $\beta$-MnO$_2$ showing the approximate octahedron formed by the oxygen (small red spheres) coordination of the central manganese (large purple spheres). First four nearest neighbor magnetic couplings also indicated. (b) The magnetic unit cell with 7/2 structural units along the c-axis. The magnetization vector is illustrated for the helical order from Yoshimori\cite{Yoshimori1}. Note the atoms are included to indicate scale, but do not represent an integer number of structural unit cells.}
  \label{fig:unitcell}
\end{figure}

The material $\beta$-MnO$_2$ is of technological and fundamental scientific interest, being the most stable and abundant polymorph of the MnO$_2$ oxide family. Technologically it
is important as a cathode material in lithium ion batteries \cite{Feng1} and as a catalyst\cite{Debart1, Lima1}. Experimentally $\beta$-MnO$_2$ occurs in the rutile structure\cite{Baur1} with space group $P4/mnm$ (\#136) and lattice parameters $a =  3.3983$ \AA~and $b=2.8730$ \AA. It orders magnetically at $T_N$ = 92 K\cite{Ohama1}. Below this temperature, transport measurements indicate an enhancement of the resistivity up to large values ($\sim$ 10$^5$ $\Omega$ cm) at $T=$0 K, indicative of insulating behavior\cite{Sato1}. The magnetic transition has been investigated in depth\cite{Yoshimori1, Regulski1, Regulski2} and results in a helical antiferromagnetic order with pitch (7/2)\textbf{c}. For simplicity, this paper initially focusses on collinear antiferromagnetic order between neighboring Mn sites along $\left\langle 111 \right\rangle$. This corresponds to opposing spins on the center and corner manganese sites in the unit cell shown in Fig.~\ref{fig:unitcell}(a). Antiferromagnetic coupling of these sites has been shown to be key to the helical magnetic order\cite{Yoshimori1, Ohama1}. Following this we return to treat the non-collinearity explicitly. In Fig.~\ref{fig:unitcell}(b) the helical order of pitch (7/2)\textbf{c} in the magnetic unit cell is shown.

Two DFT+U approaches are compared in this work. The first we term the `full anisotropy' approach, where $U$ and $J$ are both defined distinctly and the resulting Coulomb and exchange matrices respect the full orbital dependence of the \textit{d}-\textit{d} interactions. The second, introduced by Dudarev \textsl{et al.}\cite{Dudarev1}, is denoted the `effective U' approach where only $U$ is distinctly defined as $U_{\mathrm{Eff}}=U-J$, and $J$ is set to zero throughout.

It is important to note that a large majority of DFT+U calculations in the literature are performed with the effective U and therefore this work is of importance to many researchers. In some codes only the effective U approach is implemented (CASTEP\cite{Clark1}), while in others it is the default (VASP\cite{Kresse1}) or is recommended (Wien2k\cite{Blaha1}).

Previous electronic structure calculations of $\beta$-MnO$_2$ have been performed by Franchini \textit{et al.}\cite{Franchini1}. Similar to the present work, they found that the effective U method predicts a metallic ferromagnetic (FM) ground state over a wide range of $U$ values. This contravenes the experimental evidence\cite{Sato1} which is, in turn, in better accord with the insulating antiferromagnetic (AFM) ground state obtained by Franchini \textit{et al.} using hybrid functionals. It is noted, though, that the latter study did not consider the non-collinear magnetic order, but did suggest that the deficiency of the effective U approach may be due to the fact that the improved treatment of correlations is limited to localized $d$-states. In this study we show quantitatively that the inclusion of anisotropic Coulomb and exchange interactions in DFT+U corrects the effective U description.

An early periodic Hartree-Fock (HF) study by Mackrodt and Williamson\cite{Mackrodt1} found a very large gap of approximately 4 eV, the FM ground state being favored over AFM by 60 meV per formula unit. Other studies by Balachrandan \textit{et al.} have also considered $\beta$-MnO$_2$ as part of wider studies relating to polymorphic MnO$_2$ as a lithium battery insertion cathode\cite{Balachandran1, Balachandran2}. A recent DFT+dynamical mean field theory (DFT+DMFT) study\cite{Yu1} also found a gap in the excitation spectra of AFM $\beta$-MnO$_2$ within the fixed experimental structure. However, the total energies of likely magnetic states were not compared. Furthermore the magnetic order was treated within an ``appropriate cubic" cell that does not fully represent the experimental magnetic order, while the Hamiltonian employed was adapted to a cubic symmetry that $\beta$-MnO$_2$ does not possess. Our work is the first to study the energetic stability of the full non-collinear order.

Previous studies have demonstrated that anisotropic interactions in DFT+U may give rise to spin ordering of orbitals\cite{Kasinathan1} and influence the canting angle in non-collinear magnetic systems\cite{Bousquet1}. 

\section{\label{sec:dftpu}The DFT+U correction}
The DFT+U method in essence introduces a correction to traditional DFT energies based upon the occupation of correlated orbitals. Typically these orbitals correspond to strongly localized $d$ and $f$ states, wherein the on-site Coulombic and exchange interactions between electrons, usually described by $U$ and $J$, are not adequately treated by traditional functionals. We may write this correction to the total energy in the form:
\begin{align}\label{eq:PUCorrection}
	\Delta E = E_I - E_{dc}
\end{align}
The second term is a double counting correction which has been discussed in detail previously\cite{Ylvisaker1}, while the first corresponds to the interactions between the correlated electrons:
\begin{align}
	E_I = \frac{1}{2} \sum_{m \sigma \neq m' \sigma'} W_{mm'}^{\sigma \sigma'} n_{m \sigma}n_{m' \sigma'}
\end{align}
Here, for simplicity and clarity we write the sum in the Hilbert space in which the occupations are diagonal, which can always be achieved by a rotation. The Coulomb matrix $W_{mm'}^{\sigma \sigma'}$ may be expressed as:
\begin{align}
	W_{mm'}^{\sigma \sigma'} = U_{mm'} - J_{mm'} \delta_{\sigma \sigma'}
\end{align}
$U_{mm'}$ are the direct Coulomb interactions which operate between states irrespective of spin and $J_{mm'} \delta_{\sigma \sigma'}$ defines the spin-dependent exchange interactions. The elements of the two interaction matrices are all derived from the parameters $U$ and $J$ taken as input to the DFT+U method; however, what is often not appreciated is that the direct interaction $U_{mm'}$ is determined not only by $U$, but also by $J$.

It is $J$ that determines the anisotropy of the Coulomb interactions\cite{Johannes1} in the DFT+U scheme, via a linear
combination of Slater integrals. In essence these interactions have a tendency to increase the mutual distance of each pair of electrons and therefore promote Hund's second rule. Approaches to DFT+U based on an effective U omit this anisotropy by setting $J$ to zero resulting in $U_{mm'} = U$ and $J_{mm'}= 0$ for $m \neq m'$.
In contrast, with full anisotropy the Coulomb and exchange matrices (for $U=6.7$ and $J=1.2$ eV) are:
\begin{align*}
U_{mm'} \text{~~~~~~~~~~~~~} & \text{~~~~~~~~~~~~~} J_{mm'} \\
\begin{pmatrix}
7.6 & 6.2 & 5.9 & 6.2 & 7.6 \\
6.2 & 7.2 & 6.8 & 7.2 & 6.2 \\
5.9 & 6.8 & 8.1 & 6.8 & 5.9 \\
6.2 & 7.2 & 6.8 & 7.2 & 6.2 \\
7.6 & 6.2 & 5.9 & 6.2 & 7.6 
 \end{pmatrix} , &
\begin{pmatrix}
7.6 & 1.3 & 1.1 & 0.5 & 1.0  \\
1.3 & 7.2 & 0.7 & 1.9 & 0.5  \\
1.1 & 0.7 & 8.1 & 0.7 & 1.1 \\
0.5 & 1.9 & 0.7 & 7.2 & 1.3 \\
1.0 & 0.5 & 1.1 & 1.3 & 7.6
\end{pmatrix}
\end{align*} 
The off-diagonal elements of the Coulomb $U_{mm'}$ and exchange $J_{mm'}$ matrices vary significantly due to the anisotropy of the interactions within the localized $d$-manifold. 

The input parameters $U$ and $J$ are translated into matrices $U_{mm'}$ and $J_{mm'}$ by appealing to the atomic physics of the ion under HF theory\cite{Czyzyk1}. This well-established method provides an accurate description of interactions in isolated atoms\cite{Griffith1} (excepting those processes that involve a change in the number or significant change in the character of pair correlations, e.g. spin excitations) and is therefore modelled within the DFT+U approach. HF methods, including on-site hybrids\cite{Novak1}, inherently involve the computation of integrals over the occupied wavefunctions to determine the Slater integrals $F^0$, $F^2$, $F^4$ and $F^6$.  A linear combination of such integrals then determines the Coulomb and exchange matrices. However, in the DFT+U method, rather than calculating the Slater integrals explicitly we parameterize them using $U$ and $J$. Thus, for the $d$-orbital manifold of interest here,  $F^0 = U$ and, guided by atomic calculations, we set $F^2 + F^4 = 14 J$ with the condition that $F^2 / F^4 = 8/5$. Higher order integrals occur only for $f$-electrons. One advantage of such a parameterization is that we may directly apply values of $U$ and $J$ that are more representative of the effects of screening due to both correlations and interaction with the crystalline environment, with renormalized $U$ values typically in the range 4 - 8 eV, whereas a direct HF calculation of $U = F^0$ provides an unrealistic value of approximately 22-26 eV.

\section{\label{sec:electronic}Electronic structure results}
We have calculated the electronic structure using GGA+U. The VASP\cite{Kresse1} code was employed using the PAW method with a plane-wave cut-off of 600 eV. A minimum 6 $\times$ 6 $\times$ 6 \textbf{k}-point grid was used in the full Brillouin zone. For comparison, key results were also computed separately using the full potential LAPW approach in Wien2k\cite{Blaha1} with RK$_{max}$ = 7.0 and muffin tin radii of 2.01 a$_0$ for Mn and 1.51 a$_0$ for O. The fully localized limit (FLL)\cite{Ylvisaker1} double counting correction was employed which is considered approprite for localized \textit{d}-states in $\beta$-MnO$_2$.

We directly obtain $U_{\mathrm{Eff}}$ by the \textit{ab initio} method as applied in Wien2k\cite{Madsen1}, yielding a screened value $U_{\mathrm{Eff}}$ = 5.5 eV as might appropriately be applied within the spherically averaged effective U approach. Since the Slater integrals, $F^2$ and $F^4$, are typically only weakly screened in solids\cite{Antonides1, Anisimov1}, we use the atomic-limit\cite{Madsen1} value for Mn$^{4+}$ $J = 1.2$ eV. Therefore, in the full anisotropy GGA+U approach, we set $U = 6.7$ eV and $J = 1.2$ eV. In work to be published elsewhere, we show that these values yield accurate lithium intercalation voltages for $\beta$-MnO$_2$ acting as a cathode in lithium ion cells.

\begin{table}
\caption{\label{table:energies} Unit cell dimensions, unpaired spin moments and energies of antiferromagnetic (AFM) and ferromagnetic (FM) ordering for $\beta$-MnO$_2$ calculated with VASP. Experimental structural data from Regulski \textit{et al.}\cite{Regulski3}. Total energy for ferromagnetic order is set as the zero reference.}
		\begin{ruledtabular}
		\begin{tabular}{ccc||cc||c}
 &\multicolumn{2}{c||}{$U_{\mathrm{Eff}}=5.5$, $J=0$ eV}&\multicolumn{2}{c||}{$U=6.7$, $J=1.2$ eV} & Exp.\\
\cline{2-6}
 &AFM&FM&AFM&FM &\\
\hline
$a$ (\AA)& 4.462 & 4.496 & 4.443 & 4.451  & 4.398 \\
$c$ (\AA)& 2.955 & 2.991 & 2.936 & 2.940  & 2.874 \\
$\mu$ ($\mu_B$)& 3.24 & 3.61  & 2.95 & 3.09 & \\
$E$ (meV) & +48 & 0 & -20 & 0  & \\
		
\end{tabular}
\end{ruledtabular}
\end{table}

In Table~\ref{table:energies} we compare the relative total energies obtained for FM versus AFM ordering from VASP. The effective U approach gives the lowest total energy for a FM spin configuration with a stabilization of 48 meV per formula unit. In contrast, full anisotropy GGA+U shows an energetic stabilization of 20 meV per formula unit for AFM order. A qualitatively similar result is also obtained using the local spin density approximation + U (LSDA+U) method, indicating that the relevant physics is not specific to a particular choice of local or semi-local functional. We have verified in Wien2k using GGA+U that AFM is stable (by 35 meV per formula unit) only when full anisotropy is included, leading to the conclusion that this effect is not implementation or basis set dependent. However, applying the effective U approach with the LAPW method in Wien2k leads to a magnetic energy difference of just 2 meV per formula unit;  the two magnetic orders are essentially degenerate. We return to discuss this in section~\ref{sec:noncollinear} where the exchange constants dictating the magnetic ground state are calculated.

We have also tested the AFM stabilization energy for other reasonable ranges of $U_{\mathrm{Eff}}$ (4 - 6 eV) and $J$ (0.9 - 1.4 eV) and find that, while the magnitude varies, AFM order is always favored by application of the fully anisotropic approach. Most importantly, the large difference in the relative energies of FM and AFM states is robust. 

As is common with the GGA+U method, Table~\ref{table:energies} shows that the cell parameters are overestimated, but the agreement with experiment is improved when anisotropy and the lower energy AFM order are both employed. The unpaired electron count of 2.95 obtained with full anisotropy is consistent with the unpaired spin density of a Mn$^{4+}=d^3 \Rightarrow 3 \mu_B$ valence state formally expected for MnO$_2$. As discussed by Ylvisaker\cite{Ylvisaker1} \textit{et al.} the inclusion of $J$ can lead to such lower energies for smaller magnetic moments by promoting spin and orbital polarization. We note that the calculated unpaired electron count of 2.95 does overestimate the experimental\cite{Regulski3} low temperature \textit{ordered} moment of 2.4$\mu_B$. These experiments include contributions from orbital moments and spin excitations that are not captured by our calculations. Furthermore, there is a large phase space for spin fluctuations available in this system due to the large number of competing magnetic orders\cite{Yoshimori1, Blundell1}. To capture the effects of these fluctuations a theory with dynamical frequency dependent interactions may be required\cite{Petukhov1}.

\begin{figure}
\includegraphics[scale=0.6,angle=0]{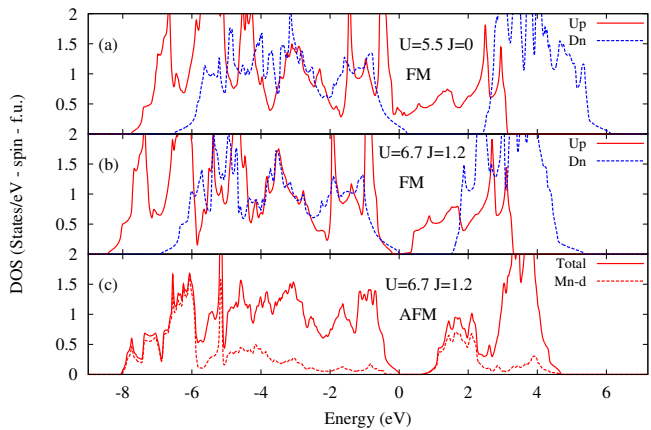}
\caption{\label{fig:DOS} (Color online) Density of states of $\beta$-MnO$_2$ from Wien2k. Panel (a) effective U ferromagnetic, (b) is full anisotropy ferromagnetic and (c) full anisotropy antiferromagnetic. (f.u.$\equiv$formula unit.).}
\end{figure}

In Fig.~\ref{fig:DOS} we show the electronic density-of-states (DOS) computed for $\beta$-MnO$_2$. The top two panels provide the DOS in FM order, Fig.~\ref{fig:DOS}(a) showing the result for the effective U approach while Fig.~\ref{fig:DOS}(b) includes anisostropy in the GGA+U treatment. The anisotropic $J$ interactions introduce a small gap while the effective U approach yields a metallic density-of-states. Fig.~\ref{fig:DOS}(c) shows the DOS for the AFM spin configuration obtained from GGA+U with full anisotropy, wherein a band gap of approximately 0.8 eV in width is evident, consistent with experimental measurements\cite{Sato1}.

\begin{figure}
\includegraphics[scale=0.7,angle=0]{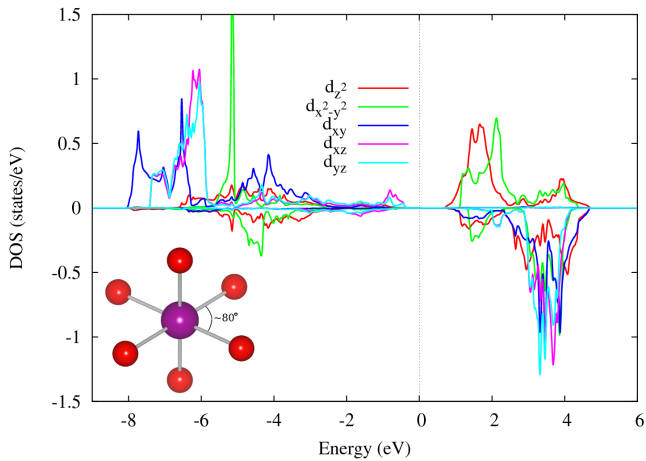}
\caption{\label{fig:LMDOS} (Color online) Projected density of states of the Mn \textit{d}-orbitals for a full anisotropy antiferromagnetic calculation from Wien2k. Postive is spin up and negative spin down. The inset illustrates the distortion of the approximate MnO$_6$ octahedron with an in-plane O-Mn-O angle of $\sim$80$^{\circ}$.}
\end{figure}

It is instructive to see what the effect on the underlying spin and charge density is due to anisotropic interactions. The Mn local environment has two apical Mn-O bonds that are at 90$^{\circ}$ to the four in-plane Mn-O bonds together comprising the approximate octahedron. However, the angle between the in-plane oxygens deviates from 90$^{\circ}$ to values of approximately 100$^{\circ}$ and 80$^{\circ}$ for consecutive pairs as shown in the inset to Fig.~\ref{fig:LMDOS}. Consequently, the octahedral symmetry is approximate and to obtain the orbital projected partial charges we set the \textbf{z}-axis along the apical Mn-O bonds and the \textbf{x,y}-axes within $\sim5^{\circ}$ from the in-plane Mn-O bonds. In Table~\ref{table:orbitalOccs} the partial charge and net spin for the Mn \textit{d}-orbitals is shown. The results indicate that the crystal field dominates the orbital occupations giving a clear $e_g$ to $t_{2g}$ splitting and Hund's first rule is obeyed with majority spin $t_{2g}$ occupations close to unity. 

The partial charges in Table~\ref{table:orbitalOccs} obtained with full anisotropy and effective U are very similar, all deviations being less than 0.01. In contrast the net spin per orbital varies significantly between the two approaches. For example the $d_{z^2}$ net spin is 0.199 for effective U and 0.090 for full anisotropy. The deviations are also large for the other $e_{g}$ orbital, $d_{x^2 - y^2}$, with a net spin of 0.183 for effective U and 0.075 for full anisotropy. However, for the $t_{2g}$ orbitals $d_{xy}$, $d_{xz}$ and $d_{yz}$ there is little difference in the spin occupations between effective U and full anisotropy. It is the net spin of the $e_{g}$ orbitals that is reduced by full anisotropy. Small changes to the spin polarization of these predominantly bonding orbitals can have a large impact on the magnetic exchange. The full anisotropy approach, by respecting the full symmetry of the \textit{d}-\textit{d} interactions gives a better description of the orbital spin polarizations.

To obtain further insight into the type of states associated with each orbital Fig.~\ref{fig:LMDOS} shows the projected density of states of the Mn \textit{d}-orbitals for a full anisotropy antiferromagnetic calculation. The large peaks in the majority up spin clustered between -8 and -5 eV indicate the position of the more localized $t_{2g}$ orbitals. There is also a sharp peak for $d_{x^2 - y^2}$ near -5 eV, but otherwise broader structure. This sharp peak is a result of the distortion of the in-plane O-Mn-O angle to $\sim$80$^{\circ}$ which results in partial charge, associated with $d_{xy}$ in the case of perfect octahedra, being attributed to $d_{x^2 - y^2}$. This has been confirmed by calculating the projected density of states for a crystal structure relaxed with the constraint that all octahedral angles remain 90$^{\circ}$ in which case the sharp $d_{x^2 - y^2}$ feature is not present. In contrast the structure of the $d_{z^2}$ partial DOS is very broad and likely to be dominated by bonding with the apical oxygen.

\begin{table}
\caption{\label{table:orbitalOccs} Projected orbital charge and net spin occupations on the Mn site for an antiferromagnetic spin configuration of $\beta$-MnO$_2$ obtained with Wien2k.}
		\begin{ruledtabular}
		\begin{tabular}{ccccc}
 & \multicolumn{2}{c}{Effective U}&\multicolumn{2}{c}{Full Anisotropy} \\
 & Charge & Spin & Charge & Spin \\
\hline
 $d_{z^2}$         & 0.609 & 0.199 & 0.599 & 0.090 \\
 $d_{x^2 - y^2}$   & 0.640 & 0.183 & 0.634 & 0.075 \\
 $d_{xy}$          & 0.965 & 0.875 & 0.965 & 0.879 \\
 $d_{xz}$          & 1.021 & 0.839 & 1.020 & 0.843 \\
 $d_{yz}$          & 1.008 & 0.849 & 1.007 & 0.854 \\
\end{tabular}
\end{ruledtabular}
\end{table}

\subsection{\label{sec:bandgap}Effect of Anisotropic Interactions on the Band Gap}
The optimized crystal structure obtained applying full anisotropy was retained and the DOS with antiferromagnetic order computed within the effective U approach, resulting in a band gap of 0.4 eV, as compared with 0.8 eV when full anisotropy is employed. Here, the use of the same crystal structure allows us to discriminate between electronic and structural effects, so that we may confidently say that the application of the effective U approach leads to a reduction in band gap of 0.4 eV. This effect can be understood within a Kanamori-type model \cite{Kanamori1} of the interactions within the localized $d$-orbital manifold, allowing us to straightforwardly rationalize the key physics occurring in the \textit{ab initio} calculations. The Hamiltonian for the $d$-states may be written:
\begin{align}
	\hat{H} = \hat{H}_{CF}^{0} + E_I
\end{align}
Here, $\hat{H}_{CF}^{0}$ incorporates the effect of the local environment (crystal field) of the cation and $E_I$ represents the DFT+U interaction term in equation~\ref{eq:PUCorrection}. Since the Mn$^{4+}$ ions occupy approximately octahedral sites, we will assume that the highest occupied $d$-state is $t_{2g}^{\alpha}$, where alpha denotes the spin index, with an excitation across the band gap into an unoccupied $e_{g}^{\alpha}$ state. Consequently, the $d$ to $d$ bandgap may be approximated as the difference in eigenvalues between these two states i.e. $E_g = \epsilon(e_{g}^{\alpha}) - \epsilon(t_{2g}^{\alpha})$.

The DFT+U potential matrix elements with the fully localized limit double counting correction have previously been expressed by Ylvisaker \textit{et al.} (see eq. 25 of ref.\cite{Ylvisaker1}):
\begin{align}
	\Delta v_{m\sigma} = -(U-J)\left[ n_{m\sigma} - \frac{1}{2} \right] + \Delta v_{m\sigma}^{\text{aniso}}
\end{align}
The first term encompasses the isotropic terms and is identical in the effective U and full anisotropy approaches. The second term $\Delta v_{m\sigma}^{\text{aniso}}$ incorporates all of the effects of anisotropic interactions in DFT+U and will be zero in the effective U case, so that the difference in band gap in the effective U and full anisotropy approaches must be largely attributable to the effect of this term alone. Below we detail the action of the latter term in our Kanamori Hamiltonian for $\beta$-MnO$_2$.

The eigenvalues of the model Hamiltonian may be evaluated using the Slater-Janak theorem as the derivative of the total energy with respect to the occupation number of the state $\epsilon(i) = \frac{dE_{Tot}}{dn_{i}}$. For the effective U case:
\begin{eqnarray}
 \epsilon(t_{2g}^{\alpha}) &=& 2U_{\text{Eff}}      \nonumber \\
 \epsilon(e_{g}^{\alpha})  &=& 3U_{\text{Eff}} + \Delta_{CF} \nonumber \\
 \text{Gap~~~~~} E_g^{\text{EffU}} &=& U_{\text{Eff}} + \Delta_{CF}
\end{eqnarray}
Here $\Delta_{CF}$ is the octahedral crystal field splitting. For full anisotropy:
\begin{eqnarray}
 \epsilon(t_{2g}^{\alpha}) &=& 2U - \frac{640}{273}J   \nonumber \\
 \epsilon(e_{g}^{\alpha})  &=& 3U - \frac{226}{91}J + \Delta_{CF} \nonumber \\
 \text{Gap~~~~~} E_g^{\text{Ani}} &=& U_{\text{Eff}} + \frac{235}{273}J + \Delta_{CF}
\end{eqnarray}

Considered together, the difference in bandgap between the two methods is $E_g^{\text{Ani}} - E_g^{\text{EffU}} = (235/273)J$, amounting to 1.03 eV with a setting $J=1.2$ eV as in the \textit{ab initio} calculations. However, we note that the band gap possesses a charge transfer character, evident from the dominant weight of oxygen $p$-states present in the upper valence bands of Fig.~\ref{fig:DOS}(c). This implies that the apparent variation in band gap between the two approaches will be due to relative motion of the upper \textit{d}-eigenvalue alone, amounting to a halving of the value above to 0.5 eV prior to comparison with the \textit{ab initio} calculations. This is broadly consistent with the variation in band gap of 0.4 eV that we see in the \textit{ab initio} calculations. The small difference is likely due to the neglect of the effects of differences in the degree of Mn-O orbital hybridization between the two approaches. This stems from the fact that the $\Delta_{CF}$ parameter above should more realistically be considered as a ligand field rather than a crystal field splitting, in that it arises from a combination of rigid ion electrostatic and hybridization effects. 

\subsection{\label{sec:noncollinear}Noncollinear Magnetism}
The full helical magnetic order with a pitch of (7/2)\textbf{c} first determined by Yoshimori\cite{Yoshimori1} is also considered here, where we assume that any effects due to the 4\% deviation from this pitch refined by Regulski \textit{et al.} are small\cite{Regulski1}. A supercell consisting of a seven unit cell expansion along the c-axis was therefore constructed, wherein the helically ordered moments lie within the \textbf{ab}-plane. Defining a propagation vector $k = (0, 0, k_z = 2/7)$, the Mn sites positioned at (0,0,0) in each of the original unit cells ($z_n = n$) take moments:
\begin{align}
	\mathbf{M}_n^{(0,0,0)} = M [\mathbf{sin}(2\pi k_z z_n), \mathbf{cos}(2\pi k_z z_n),0]
\end{align}
where $n$ indexes the units in the supercell. Meanwhile, the ($\frac{1}{2},\frac{1}{2},\frac{1}{2}$) positions ($z_n = n + 1/2$) in each unit cell take moments:
\begin{align}
	\mathbf{M}_n^{(\frac{1}{2},\frac{1}{2},\frac{1}{2})} = M [-\mathbf{sin}(2\pi k_z z_n), -\mathbf{cos}(2\pi k_z z_n),0]
\end{align}
The total energy of this full helical order relative to FM ordering has been computed using the non-collinear magnetism approach within the VASP code. Application of the effective U method leads to the stabilization of FM order by 65 meV per formula unit, while the inclusion of anisotropic interactions, with $U$ and $J$ as defined above, results in a stabilization of the full helical order by 10 meV per formula unit. Inspection of the DOS for these non-collinear calculations gives band gaps that differ by at most 0.1 eV from those shown in Fig.~\ref{fig:DOS} for the collinear calculations. This is due to the low energy scale for the non-collinearity which results in no qualitative difference to our conclusions regarding the band gap.

\begin{table*}
\caption{\label{table:Heis} Exchange parameters in Kelvin obtained by fitting equation~\ref{eq:Heis} to total energies calculated for a range of magnetic states. Experimental results obtained by Ohama \textit{et al.} are included for comparison\cite{Ohama1}. Derived parameters $J_{1NN}/J_{2NN}$ and $2J_{1NN}J_{3NN}/J_{2NN}^2$ are not shown for the VASP effective U results on the basis that $J_{1NN}$ and $J_{2NN}$ do not satisfy Yoshimori's conditions for a stable helical state\cite{Yoshimori1}.}
		\begin{ruledtabular}
		\begin{tabular}{cccccccc}
		& \multicolumn{2}{c}{Wien2k LAPW} & \multicolumn{2}{c}{VASP PAW} & \multicolumn{2}{c}{CRYSTAL09} & Experiment \\
 & Effective U & Full Anisotropy & Effective U & Full Anisotropy & B3LYP & PBE0 & \\
\hline
$J_{1NN}$ (\textbf{K}) & 4.3$\pm$0.9 & 9.0$\pm$1.7 & -19.1$\pm$6.2  &  4.3$\pm$0.7 & 13.2$\pm$2.1 & 8.1$\pm$1.5 & 8.9 \\
$J_{2NN}$ (\textbf{K}) & 2.2$\pm$0.3  & 6.4$\pm$0.6 & -11.5$\pm$2.1 &  3.6$\pm$0.3 & 8.3$\pm$0.7 & 5.8$\pm$0.5 & 5.5 \\
$J_{3NN}$ (\textbf{K}) & 2.4$\pm$0.6  & 2.4$\pm$1.2 & 0.0$\pm$4.3   &   1.8$\pm$0.6 & 4.0$\pm$1.4 & 3.2$\pm$1.0 & -1.3 \\
$J_{4NN}$ (\textbf{K}) & -0.9$\pm$1.5  & -0.4$\pm$0.6 & -1.2$\pm$2.2 & -0.5$\pm$0.3 & -0.3$\pm$0.5 & -0.4$\pm$0.6 & - \\
\hline
$J_{1NN}/J_{2NN}$  &  1.95$\pm$0.48  &  1.41$\pm$0.29 & - &  1.19$\pm$0.21 & 1.59$\pm$0.29 & 1.40$\pm$0.28 & 1.6\footnote{The ratio 1.6 was assumed in fitting to the neutron time of flight experiments.} \\
$2J_{1NN}J_{3NN}/J_{2NN}^2$  &  4.26$\pm$1.62  &  1.05$\pm$0.58  & - &  1.19$\pm$0.46 & 1.53$\pm$0.62 & 1.54$\pm$0.59 & -0.76 \\
\end{tabular}
\end{ruledtabular}
\end{table*}

Yoshimori\cite{Yoshimori1} has studied the magnetic structure of $\beta$-MnO$_2$ using a classical Heisenberg spin Hamiltonian including the first three Mn-Mn neighboring interactions $J_{1NN}$, $J_{2NN}$ and $J_{3NN}$ as shown in Fig.~\ref{fig:unitcell}(a). The helical magnetic order observed experimentally was found to be stable if: (1) $J_{1NN}$, $J_{2NN}$ and $J_{3NN}$ are all positive (favoring AFM order), (2) $J_{1NN} > J_{2NN}$ and (3) $J_{3NN}$ is small or negative compared to $J_{2NN}$ according to the condition $2J_{1NN}J_{3NN}/J_{2NN}^2 < 1$. To compare our DFT results to these conditions we have chosen to map the DFT energies onto a classical Heisenberg Hamiltonian\cite{Middlemiss1, Feng2}:
\begin{align}\label{eq:Heis}
	\hat{H} = E_0 + 2\sum_{a}\sum_{\langle i,j \rangle} J_{i,j}^aS_i^a \cdot S_j^a
\end{align}
where $a$ labels progressively more distant shells of Mn-Mn interactions. Couplings out to fourth-nearest-neighbor are explicitly considered here, but, as will be shown below, the fourth neighbor interactions are small enough that they may reasonably be neglected. The second summation runs over distinct spin pairs occurring at each appropriate shell separation, $a$, where $S_i^a$ is the three dimensional spin vector on site $i$ and $E_0$ is a constant term that can be thought of as the non-magnetic energy of the system. For example, the energy of FM order within this Hamiltonian is obtained as $E_{FM}=E_0+[2J_{1NN}+8J_{2NN}+4J_{3NN}+8J_{4NN}]S^2$. The Heisenberg model was least-squares fitted to the total energies of 13 different Ising-type spin configurations representable within a $2\times2\times2$ supercell expansion of the appropriate relaxed crystal structure presented in Table~\ref{table:energies}. To be clear, Ising-type configurations restrict the spin pairs occurring above to be fully co- or anti-aligned.   The results from Wien2k and VASP for both effective U and full anisotropy are presented in Table~\ref{table:Heis}. The table includes standard errors for all parameters obtained from the fit and the magnitude of the spin vector is $S=3/2$, consistent with previous work and the $d^3$ formal configuration. Experimental results from the time-of-flight neutron study by Ohama \textit{et al.} are included for comparison\cite{Ohama1} where a 15 \% error was inferred from the fit to the experimental data.

Examining first the results from Wien2k, the effective U calculation yields AFM couplings for the first three nearest neighbor couplings. However, both $J_{1NN}$ and $J_{2NN}$ amount to less than half of the values obtained experimentally. Furthermore, while the effective U calculation satisfies Yoshimori's second condition for a helical state with $J_{1NN}/J_{2NN} = 1.95$ it grossly fails the third condition with a third nearest neighbor coupling that is too large resulting in $2J_{1NN}J_{3NN}/J_{2NN}^2 = 4.26$. Even within the range of error on this parameter, $\pm$1.62, this places the spin system far from the stability condition for the helical state. Meanwhile, using Wien2k AFM couplings are again obtained from the full anisotropy calculation for the first three neighbors, but are significantly larger in magnitude. The nearest two neighbor couplings deviate by less than 17\% from those obtained experimentally and satisfy Yoshimori's second condition for a helical state with $J_{1NN}/J_{2NN} = 1.41$. The third condition is borderline at $2J_{1NN}J_{3NN}/J_{2NN}^2 = 1.05$, but the large error renders it difficult to make a definitive statement as to the smallness of $J_{3NN}$. Clearly, the inclusion of full anisotropy improves the description of the spin system by increasing the magnitude of $J_{1NN}$ and $J_{2NN}$, as well as bringing the system to the border of the third stability condition for helical order. The corresponding mean field transition temperature may be calculated as $T_N=\left\{S(S+1)/3k_B\right\}(8J_{2NN}^2/J_{1NN}+4J_{1NN}-8J_{3NN})\approx 66$ \textbf{K}, in reasonable agreement with the experimental transition of 92 \textbf{K}. Another important finding is that the fourth nearest neighbor coupling $J_{4NN}$ is small in all calculations, and always occurs with a large fractional error, suggesting that its contribution to the magnetic energy is ill defined.

Turning to examine the VASP results, the effective U calculation with PAW potentials produces FM couplings for the first two nearest neighbors and therefore clearly fails Yoshimori's first and most basic condition determining the stability of helical order. It is interesting that the effective U calculation using full-potential LAPW in Wien2k, by comparison, retains AFM couplings, albeit that they are small. This demonstrates that the approximations made to the potential in the PAW approach adversely affect the prediction of exchange constants in this system. A similar effect likely also underlies the difference in magnitude of the energetic stabilities found for collinear order with effective U in VASP and Wien2k. The full anisotropy calculation with VASP recovers AFM couplings for the three closest neighbors, although they are still significantly smaller than those found experimentally or those with full anisotropy in Wien2k, resulting in a low mean field transition temperature $T_N= 35$ \textbf{K}. Therefore, it seems clear that the inclusion of full anisotropy favors AFM coupling, but the apparent underlying bias of the PAW potentials toward FM coupling results in net smaller couplings in VASP as compared with Wien2k. The calculations in both Wien2k and VASP are performed in the rotationally invariant scheme. The reasons for the poorer couplings from VASP warrants future work and may include not only the method for selecting the orbitals on which $U$ acts, but also the details of the pseudization in the PAW potentials, the interplay between the
 Hubbard-like Hamiltonian and the pseudo-projectors and the use of a second variational procedure for the incorporation of the DFT+U potential.

At this point it is timely to make some further general comments regarding the fitting of the exchange couplings. Firstly, theory in all cases predicts a positive (AFM) $J_{3NN}$, while the experimental values yield a small negative (FM) coupling $J_{3NN}=-1.3$ \textbf{K}. However, it must be emphasised that the experimental results were obtained enforcing a ratio $J_{1NN}/J_{2NN} = 1.6$ in the fitting, which has an unknown effect on the $J_{3NN}$ value. Secondly, four-parameter fits in which $J_{4NN}=0$ is enforced showed slightly decreased values of $J_{1NN}$ and $J_{3NN}$ with little change in $J_{2NN}$. For example, the full anisotropy calculation in Wien2k then gives $J_{1NN}=8.4 \pm 1.3$ \textbf{K}, $J_{2NN}=6.3 \pm 0.5$ \textbf{K} and $J_{3NN}=1.9 \pm 0.8$ \textbf{K}. In systems possessing rich magnetic phase diagrams, such as the present $\beta$-MnO$_2$, variations of this magnitude could conceivably alter the nature of the predicted low temperature ground state, in particular here whether the stability conditions set out by Yoshimori for helical order are satisfied. However, such small changes in coupling do not alter the overriding conclusion that fully anisotropic interactions improve the description of the exchange constants.

It may also be suggested that the description of the magnetism and band gap could be improved by the use of a smaller value of $U$ within the effective U approach. Indeed the antiferromagnetic band gap only closes for values of $U_{\mathrm{Eff}}$ below 3 eV. Employing $U_{\mathrm{Eff}}=3.5$ eV one may stabilize the \textit{collinear} antiferromagnetic order by very small energies while still retaining a band gap. However, a full fitting of the exchange integrals in Wien2k with $U_{\mathrm{Eff}}=3.5$ eV results in $J_{1NN}=8.5 \pm 1.1$, $J_{2NN}=4.0 \pm 0.4$ and $J_{3NN}=2.6 \pm 0.8$. With these values $J_{1NN}/J_{2NN} = 2.12 \pm 0.37$ and $2J_{1NN}J_{3NN}/J_{2NN}^2 = 2.75 \pm 0.92$ clearly demonstrating that Yoshimori's stability conditions are not satisfied. Therefore, reducing the value of $U_{\mathrm{Eff}}$ does not remedy the poor description of the exchange couplings. Furthermore, such a small value of $U$ is not consistent with our self-consistent calculation of the parameter.

Finally, for comparison, in Table~\ref{table:Heis} the results obtained from two types of hybrid functional calculations within the CRYSTAL09 LCAO code\cite{Dovesi1, *Dovesi2} are also presented. Here, the crystal structure obtained from hybrid HSE06\cite{Heyd2} in VASP is used  giving $\mathbf{a}=4.383$ \AA~and $\mathbf{c}=2.875$ \AA~in good accord with experiment. The well-established 8-6411d41 and 6-31d1 basis sets were employed for Mn and O respectively\cite{CrystalBasis}, other numerical parameters being similar to the VASP and Wien2k calculations. The results of both functionals satisfy Yoshimori's first and second condition for the helical order, although B3LYP (bearing 20\% HF exchange)\cite{Becke1} does overestimate the first three nearest neighbour couplings compared to experiment. PBE0 (25\% HF exchange)\cite{Adamo1} gives a similar level of agreement as was obtained with full anisotropy in Wien2k. For both B3LYP and PBE0, the computed ratios $2J_{1NN}J_{3NN}/J_{2NN}^2= 1.53$ and $1.54$, respectively, appear to be too large to satisfy Yoshimori's third condition, but, again, the large errors hinder a definitive statement. 

The good performance of these hybrid functionals is consistent with the improvement obtained by use of fully anisotropic DFT+U over the effective U approach. This is not surprising, given that the proper anisotropy of \textit{d}-\textit{d} exchange interactions is included by construction in both hybrids, and that these are precisely the interactions described by full anisotropic DFT+U that do not occur in the effective U approach. Viewed more broadly, this suggests that, if the results obtained from hybrid functionals are to be compared with those from DFT+U, the latter calculations should apply fully anisotropic \textit{d}-\textit{d} interactions.


We digress to discuss some qualitative features of the computed exchange couplings, particularly the antiferromagnetic $J_{1NN}$ interaction. Antiferromagnetic coupling is predicted by all of our calculations (except effective U with PAW potentials) and is consistent with the experimental findings. However, the Mn-O-Mn bond angle of the $J_{1NN}$ interaction, at $\sim$101$^{\circ}$, is close to 90$^{\circ}$, and might therefore be predicted to yield weakly ferromagnetic coupling based upon a straightforward application of the Goodenough-Kanamori rule\cite{Goodenough1} for 90$^{\circ}$ superexchange. Beyond this simple model, the situation in the real crystal is made more complex by the further note of Goodenough that ``the cation-cation interactions tend to dominate the 90$^{\circ}$ superexchange in oxides if the t$_{2g}$ orbitals are half-filled''. Thus, direct exchange should be considered in addition to the oxygen mediated superexchange, where the former may be ferromagnetic or antiferromagnetic depending upon the relative weight of the potential and kinetic contributions (varying sensitively with the orthogonality, separation and radius of the orbitals involved). In light of the above, we suggest that an electron hopping driven antiferromagnetic direct exchange dominates the $J_{1NN}$ coupling, as supported by additional calculations wherein the c-axis length of $\beta$-MnO$_2$ is reduced such that the Mn-O-Mn angle closes to 90$^{\circ}$. In that case $J_{1NN}$ becomes even more antiferromagnetic at 139 \textbf{K} using full anisotropy GGA+U, in keeping with an antiferromagnetic direct exchange that tends to be progressively strengthened by a reduction in cation separation. By way of contrast, the 180$^{\circ}$ Goodenough-Kanamori rules are based on lower order perturbation theory and might therefore be expected to be more robust to changes in geometry, in keeping with our finding of antiferromagnetic $J_{2NN}$ couplings mediated by $\sim$130$^{\circ}$ Mn-O-Mn bridges.

Finally, Geertsma and Khomskii\cite{Geertsma1} have also suggested that the position of a ``pendant'' ligand neigboring the bridging oxygen anions may induce antiferromagnetic 90$^{\circ}$ coupling. They considered CuGeO$_3$, where the Ge cations are pendant to the Cu-O-Cu bridges formed by CuO$_2$ ribbons with nearly square planar anion coordination of the Cu$^{2+}$ cations. Meanwhile, in the present $\beta$-MnO$_2$ phase, the Mn-O-Mn bridges are formed by columns of MnO$_6$ polyhedra edge sharing along the c-axis, while the pendant species comprise the Mn cations in neighboring chains. We have quantitatively tested the effect of varying the pendant Mn to bridging oxygen distance while maintaining the bond lengths and angles of the Mn-O-Mn $J_{1NN}$ bridges at constant values. We find that a reduction in this distance of 0.1 \AA~in fact decreases $J_{1NN}$ in magnitude by 4 \textbf{K} using full anisotropy GGA+U, which is contrary to the trend expected if the pendant cations were promoting antiferromagnetic exchange. This reinforces the suggestion above that the antiferromagnetic sign of $J_{1NN}$ is largely the result of direct Mn-Mn exchange. We stress again that the majority of our computed results are consistent with experimental findings for $\beta$-MnO$_2$.

\section{Conclusions}
In summary, the primary outcome from this work is that the inclusion of fully anisotropic interactions in DFT+U is key to understanding the magnetic order and resulting band gap of $\beta$-MnO$_2$. This allows us to pinpoint the origin of the low temperature magnetic transition and resulting poor electrical conductivity that is observed experimentally. The isotropic component of \textit{d}-\textit{d} interactions is typically not treated adequately by GGA or LSDA, and there is no reason to expect that the treatment of anisotropic interactions should be any better. Therefore, the incorporation into DFT+U of fully anisotropic Coulomb and exchange interactions should be routinely considered for most \textit{d}-electron systems, particularly in light of the fact that the more accurate approach may be deployed with little additional computational effort in the majority of electronic structure codes.  More generally, these results are expected to be of importance at the interface of DFT+U methods with other treatments of correlated electrons beyond pure DFT, such as hybrid functionals, in which the anisotropy of such interactions is included by construction, and DFT+DMFT, where anisotropic interactions enter via a model Hamiltonian. 

The mechanism leading to an increase in band gap due to the presence of anisotropic interactions has been analytically demonstrated by considering the eigenvalues of a model Kanamori Hamiltonian. The Mn \textit{d}-orbital partial charges suggest that the improved treatment of full anisotropy is driven by a reduction in unphysical spin polarization of predominantly bonding orbitals. Exchange couplings have been obtained by the fitting of a Heisenberg spin Hamiltonian to the total energies of a range of magnetic states, and are found to be adversely affected by the pseudization in PAW potentials compared with full-potential LAPW and all-electron LCAO hybrid functional results. The exchange couplings obtained with DFT+U including full anisotropy and those from hybrid functionals produce the closest agreement with experimental results. The proximity of these calculated results to the threshold of the Yoshimori stability conditions for the helical order in $\beta$-MnO$_2$ suggest that this system is close to the border of competing magnetic instabilities.

\begin{acknowledgments}
We acknowledge discussions with W.E. Pickett and financial support of the EPSRC, UK and SUPERGEN Energy Storage Consortium (funded by EPSRC EP/H019596/1) as well as HECToR computer resources through the Materials Chemistry Consortium as funded by the EPSRC (EP/F067496). Research carried out (in whole or in part) on the NANO computer cluster at the Center for Functional Nanomaterials, Brookhaven National Laboratory, USA, which is supported by the U.S. Department of Energy, Office of Basic Energy Sciences under Contract No. DE-AC02-98CH10886.
\end{acknowledgments}


\bibliography{bMnO2}

\end{document}